\documentstyle[12pt]{article}
\begin{document}
\makeatletter
\@addtoreset{equation}{section}
\makeatother
\renewcommand{\theequation}{\thesection.\arabic{equation}}
\baselineskip 15pt

\title{\bf Decoherent Histories and Realism.}

\author{Angelo Bassi\footnote{e-mail: bassi@ts.infn.it}\\
{\small Department of Theoretical Physics of the University of Trieste,}\\
and \\
\\ GianCarlo Ghirardi\footnote{e-mail: ghirardi@ts.infn.it}\\
{\small Department of Theoretical Physics of the University of Trieste, and}\\
{\small the Abdus Salam International Centre for Theoretical Physics,
Trieste, Italy.}}

\date{}

\maketitle

\begin{abstract}
In this paper we reconsider the Decoherent Histories approach
to Quantum Mechanics and we analyze some problems related to its interpretation 
which,
according to us, have not been adequately clarified by its proponents.  We put 
forward
some assumptions which, in our opinion, are necessary for a realistic 
interpretation of
the probabilities that the formalism attaches to decoherent histories. We prove 
that
such assumptions, unless one limits the set of the decoherent families which can 
be
taken into account, lead to a logical contradiction.
The line of reasoning we will follow is conceptually different from other
arguments which
have been presented and which have been rejected by the supporters of the
Decoherent Histories approach.
 The conclusion is that the Decoherent Histories approach, to be considered as 
an
interesting realistic alternative to the orthodox interpretation
of Quantum Mechanics, requires the identification of  a
mathematically precise criterion to characterize an appropriate
set of decoherent families which does not give rise to any
problem.
\end{abstract}

%-------------------------- Definitions --------------------------------------
\def\tz{{t_{0}}}
\def\tu{{t_{1}}}
\def\td{{t_{2}}}
\def\tt{{t_{3}}}
\def\tn{{t_{n}}}
\def\tm{{t_{m}}}
\def\tk{{t_{k}}}
\def\pa{{P_{\alpha}}}
\def\pb{{P_{\beta}}}
\def\pg{{P_{\gamma}}}
\def\pap{{P_{\alpha'}}}
\def\pbp{{P_{\beta'}}}
\def\pgp{{P_{\gamma'}}}
\def\pau{{P^{(1)}_{k}}}
\def\pad{{P^{(2)}_{k}}}
\def\pan{{P^{(n)}_{k}}}
\def\pam{{P^{(m)}_{k}}}
\def\pak{{P^{(k)}_{k}}}
\def\qau{{Q^{(1)}_{\alpha_{1}}}}
\def\qad{{Q^{(2)}_{\alpha_{2}}}}
\def\qan{{Q^{(n)}_{\alpha_{n}}}}
\def\qam{{Q^{(m)}_{\alpha_{m}}}}
\def\qak{{Q^{(k)}_{\alpha_{k}}}}
\def\pim{{\pi^{k}_{\alpha_{m}}}}
\def\panmu{{P^{(n-1)}_{j}}}
\def\qanmu{{Q^{(n-1)}_{\alpha_{n-1}}}}
\def\paub{{P^{(1)}_{\bar{k}}}}
\def\padb{{P^{(2)}_{\bar{k}}}}
\def\panb{{P^{(n)}_{\bar{k}}}}
\def\pamb{{P^{\bar{\alpha}_{m}}_{m}}}
\def\ptan{{P^{(n)}_{\tilde{k}}}}
\def\ptanmu{{P^{(n-1)}_{\tilde{j}}}}
\def\pbm{{P^{(m)}_{j}}}
\def\his{{\textsc{His}}}
\def\fam{{\textsc{Fam}}}
\def\tr{{\textsc{Tr}}}
\def\re{{\textsc{Re}}}
\def\hisa{{\textsc{His}^{(\alpha)}}}
\def\hisb{{\textsc{His}^{(\beta)}}}
\def\hisg{{\textsc{His}^{(\gamma)}}}
\def\hisap{{\textsc{His}^{(\alpha')}}}
\def\hisbp{{\textsc{His}^{(\beta')}}}
\def\hisgp{{\textsc{His}^{(\gamma')}}}
\def\hisI{{\textsc{HisI}}}
\def\his0{{\textsc{His0}}}
\def\hisba{{\textsc{His}^{(\bar{\alpha})}}}
\def\His{{\textsc{His}}}
\def\dhis{{\textsc{D-His}}}
\def\htu{{h^{t}_{1}}}
\def\sqx{{\Sigma^{2}_{x}}}
\def\sqy{{\Sigma^{2}_{y}}}
\def\sqz{{\Sigma^{2}_{z}}}
\def\sqxp{{\Sigma^{2}_{x'}}}
\def\sqyp{{\Sigma^{2}_{y'}}}
\def\sqzp{{\Sigma^{2}_{z'}}}
%-----------------------------------------------------------------------------

\section{Introduction.}

    After more than 70 years of debate about the difficulties that
one encounters in working out a coherent view of physical
processes based on the standard formulation of quantum mechanics,
there is now a widespread belief that such difficulties do not
arise from philosophical prejudices (as has been repeatedly
asserted by many of the supporters of {\it textbook quantum
mechanics}) but represent precise mathematical and physical
challenges which call for a physical solution. As J. Bell
appropriately stated \cite{ref1} {\it the way ahead is unromantic
in that it requires mathematical work by theoretical physicists,
rather than interpretations by philosophers}. It is also
encouraging, for those who share this position, to see that now
there are explicit proposals indicating possible ways out in which
the process of measurement (and more generally all those
measurement--like processes \cite{ref2} {\it we are obliged to
admit ... are going on more or less all the time, more or less
everywhere}) is analyzed not in terms of vague assertions or
resorting to ill--defined and/or contradictory dualistic evolution
processes (the linear and deterministic evolution for microsystems
and the nonlinear and stochastic wave packet reduction for the
macroscopic ones) but in terms of more fundamental physical
concepts. In particular, approaches of this kind, which S.
Goldstein \cite{ref3} has appropriately denoted as {\it
observer--free formulations of quantum mechanics}, can be grouped
in three categories: the Pilot--Wave Theories, the Decoherent
Histories and the Spontaneous Localization models.

    This paper is devoted to a critical analysis of the so called
Decoherent Histories (DH) approach [4-14]. The motivation of such an
approach is one we completely share, i.e. to transform the probabilistic
statements of quantum theory which, in its standard formulation, are
not referring to properties actually possessed by individual physical
systems but to the potentialities of getting certain outcomes conditional
under the measurement being performed, into statements about
sequences (histories) of objective events. To each history the DH
approach attaches a probability which coincides with the one that
standard quantum mechanics attaches to a process in which the
measurements aimed to ascertain the claimed properties are actually
performed and give the stated results.

    The DH approach faces various problems, the most crucial
deriving from quantum interference effects, which render problematic
the coarse--graining of histories. The way out is obtained
by assigning probabilities not to all
conceivable histories, but only to an appropriately selected subset of
them. Actually the procedure consists in adopting a precise
mathematical criterion, the decoherence condition, which allows the
identification of families of {\it decoherent histories}, closed under
coarse--graining. Only to histories belonging to such families the formalism
attaches definite probabilities.

    Even when this move is done, the proposed DH interpretation of
the probabilities meets two serious difficulties. The first one originates
from the fact that the decoherence condition, in spite of the fact that it
limits significantly the set of the acceptable families, turns out not to be
sufficiently restrictive. For instance, Dowker and Kent \cite{ref14}
have proved that within the theory, {\it taking the past
and present for granted}, one can identify infinitely many decoherent
families of histories which imply, in general, a future non-classical
behaviour for macroscopic systems: that is, the theory has a very weak
predictive power. Quite in general, the very existence
of {\it physically senseless decoherent families}\footnote{i.e.
families whose histories  are manifestly unacceptable on physical
grounds.}  raises serious
problems of interpretation. Which meaning whatsoever can one attach to the
histories of a decoherent family claiming that the celebrated Schr\"{o}dinger's
cat is not {\it {\bf either} alive {\bf or} dead, but alive+dead}?

    The second problem is related to the occurrence of incompatible
decoherent families\footnote{Two decoherent
families are said to be incompatible when one cannot combine them
in a larger decoherent family. According to the very spirit of the DH
approach,
inconsistencies deriving from comparing different histories belonging to
incompatible families
are irrelevant: just as one cannot compare statements referring to
noncommuting
observables in standard quantum mechanics, one cannot derive conclusions from
arguments requiring the consideration of different histories belonging to
incompatible
families.}. This fact raises serious problems of interpretation whose analysis
represents a relevant part of the present paper. For the moment we confine
ourselves to recall that, according to Griffiths [4--8], the correct way to
circumvent the difficulties consists in stating that all
reasonings, all conclusions about properties possessed by physical systems,
hold only
when they are drawn within a {\it\bf single family} of decoherent
histories, or, at
most, within compatible families (we will denote this prescription as `the
single
family rule'). Otherwise, any assertion is devoid of any meaning
whatsoever. This fact
is rather puzzling and gives rise to problems with classical logic, as
stressed,
e.g., by d'Espagnat
\cite{ref15}. Actually, Griffiths [7,8]
and Omn\`es [10,11] themselves have felt the necessity, to face this
problem, to
make more precise the logical and interpretative bases of the theory.

    The present paper consists of two parts and is organised in the
following way:
the first part is devoted to present a general sketchy view of the DH approach
aimed to focus its most relevant features which will be the subject of the
subsequent discussion. In Section 2 we recall the formal aspects of the theory
with particular reference to the crucial role played by the decoherence
condition and to its physical meaning. We will also resort  to an
elementary example
to better clarify this point. In  Section 3, following the line of thought of
Griffiths
\cite{ref6} and Omn\`es \cite{ref9}, we tackle the problem of making more
explicit the logical structure of the theory by equipping the histories of a
decoherent family with the structure of a Boolean algebra. This step will
clarify the sense in which the scheme should allow to recover classical
logic and reasoning. Particular attention will be devoted (taking into
account the
fundamentally probabilistic structure of the theory) to the formalization
of the idea
 of `logical implication' between histories, a crucial step to
work out a sensible `quantum reasoning'. We will also show, by
resorting once more to an elementary example, how, to avoid inconsistencies,
one has to accept `the single family rule'. In the final section of this
first part of
the paper we  raise the problem of how many decoherent families have to
be taken
into account and we  outline the different positions of different
supporters of the DH approach about this problem.

In Part 2 of the paper we  deepen our analysis by raising the
fundamental (in our opinion) question of the nature of the
probabilities of the DH approach. In particular, in Section 4 we
discuss whether one can attach truth values to the probabilistic
statements of the theory. By resorting to elementary examples
taken from Classical and Standard Quantum mechanics we  try to
make clear that only a precise position about this crucial problem
allows to really overcome the conceptual difficulties of the
orthodox interpretation of quantum mechanics. In fact, the
question of whether the histories of a decoherent family have
truth values has a direct relation to the question of whether the
statements of the theory refer to properties of individual
physical systems which can be considered as objectively possessed
by them.  The mathematically natural way to implement the idea of
the histories possessing truth values is that of assuming that
there exist an appropriate homomorphism between the histories and
the set \{0,1\}, the two values corresponding to the falsity and
the truth of a given history, i.e., of the statements it makes
about properties possessed (in general at different times) by an
individual physical system. By resorting once more to elementary
physical examples, we stress, first of all, that truth values can
be attached only to  histories belonging to at least one
decoherent family. Thus, we will consider a precise homomorphic
map of the members of any decoherent family on the set \{0,1\},
the map being, {\it a priori}, family dependent.

This approach will allow us to discuss   the relations existing
between the homomorphism we are interested in and the `single
family rule' analyzed in Part 1 of the paper. Since the proponents
of the DH approach have not committed themselves about the truth
values of the histories, no immediate connection can be
established between such a rule and the basic features of the
homomorphism we are interested in  (contrary to the opinion of
Griffiths --- private correspondence --- and also of the referee
of this paper). Accordingly, we will investigate critically the
problem of truth assignments to the histories of any given
decoherent family.

At this stage a new problem arises: the {\it same history} (whose
probability is family independent) may belong to different and
incompatible decoherent families\footnote{Actually, given any
history belonging to at least one decoherent family there are
always other incompatible decoherent families containing it.}.
This leads naturally to raise the question of whether the
homomorphisms (one for each decoherent family) we have been led to
consider in order to give a `classical' status to the
probabilistic assertions of the theory might assign different
truth values (depending on the decoherent family to which it is
considered to belong) to the same history. We argue that this
cannot be the case unless one is keen to spoil the theory of all
its most appealing features. Accordingly, we  put forward the
assumption that the truth value which is attached to a specific
history belonging to a decoherent family is independent from the
particular family one has in mind. We stress that nothing in the
original formulation of the `single family rule' implies that such
an assumption is illegitimate for the  simple reason that the
proponents of the DH approach did not face the problem of the
truth values of the histories. However, if somebody feels that
this requirement violates the single family rule (we stress that
we have deliberately chosen to use the term `feels' because in the
formulation of such a rule nothing implies --- syntactically or
semantically --- the exclusion of such an assumption) then we are
ready to plainly declare that we are violating this particular
reading of the rule. But we will  make clear that the theory, if
one does not equip it with an homomorphism making true or false
its statements or if one accepts that the same history can have
different truth values, does not exhibit any conceptual advantage
with respect to the standard interpretation of quantum mechanics.
Then, we prove a general theorem, i.e., that the assumptions:
\begin{enumerate}
\item[a)] Every decoherent family can be equipped with a natural Boolean
structure reflecting the Boolean structure of classical logic,

\item[b)] Every history belonging to a decoherent family has a precise
truth value,

\item[c)] The truth value of any precise single  history
does not depend on the
decoherent family to which it  belongs,

\item[d)] All decoherent families have the same
status and must be taken into account,
\end{enumerate}  lead to a contradiction.

The last Section is devoted to a brief review of the main
interpretative problems analyzed in this paper. In our opinion the
most appropriate choice consists in limiting the class of all
decoherent families which can be taken into account; a similar
conclusion was reached by the authors of refs. \cite{ref17} and
\cite{ref3}. We call attention to the fact that even this program
is not easy to implement, but we leave open the question of
whether one can reach this goal in a satisfactory way. We will
reconsider the problem of further limiting the acceptable families
in a subsequent paper. \vspace{1cm}
\begin{center}
{\Large\bf  Part I}\\
\end{center}

\section{The Decoherent Histories.}

    In this section we present a concise summary of the Decoherent
Histories approach, and we show, by resorting to a simple example,
how one can  argue about the properties and about the time evolution of
quantum
systems within such a scheme.

\subsection{The formalism.}

    We consider an individual physical system $S$ --- a particle, a
macroscopic object, the whole universe --- whose initial state at time
$t_{0}$
is described by the statistical operator $W(t_{0}) = W$.
Let ${\mathcal H}$  be the Hilbert
space of the system and $U(t, t')$  the unitary operator describing its
evolution.

    We also consider an ordered sequence of time instants,
($\tu ,\td ,...,\tn$), $\tm < t_{m+1}$, $(m = 1, 2, ..., n-1)$ and,
associated to any time  $\tm$
of the sequence, an exhaustive and exclusive set\footnote{It is important
to keep in mind that the families of projection operators associated to
different times are, in general, different from each other.}
of projection operators $\{\pam\}$:
\begin{equation} \label{2.1}
    \sum_{k}\pam = 1, \qquad\quad \pam\pbm = \delta_{k, j}\pam.
\end{equation}
Besides these operators we will take also into account
all their possible sums:
\begin{equation}  \label{2.1a}
    \qam =\sum_{k} \pim \pam,
\end{equation}
where $\pim$ takes the values $0$ or $1$. Obviously,
when all possible values of $\pim$ are considered,
one gets $2^{N}$ different projection operators $\qam$ for each time
$t_{m}$, where $N$
represents the number
of the projection operators $\pam$ of the exhaustive and exclusive set
(\ref{2.1}).
\begin{description}
\item[Definition.] A {\it History} $\hisa$ is a sequence of $n$ pairs $(\qam,
\tm)$, $(m = 1, 2, \ldots, n)$
each consisting of one projection operator from the set $\{\qam\}$
and the corresponding time $\tm$:
\begin{equation} \label{2.2}
    \hisa = \left\{ (\qau, \tu), (\qad, \td), \ldots, (\qan, \tn)\right\},
\end{equation}
$\alpha$ being a shorthand for the definite sequence $(\alpha_{1}, \alpha_{2},
\ldots, \alpha_{n})$.
\end{description}
Obviously, the history $\hisa$ is assumed to correspond to the
statement: the physical system $S$ possesses, at time $\tu$, the
properties described by the projection operator $\qau$, at time  $\td$  those
described by $\qad$, and so on. When the projection operators
appearing in equation (\ref{2.2}) belong to the basic sets $\{\pam\}$, $(m
= 1, 2, \ldots, n)$, the
associated $\hisa$ are usually referred to as {\it  fine--grained
histories}, otherwise they are called {\it coarse--grained histories}.
\begin{description}
\item[Definition.] A {\it Family of Histories} $\fam$
is the set of all histories of the form (\ref{2.2}), when the projection
operators
appearing in it run over all possible members of the  sets $\{\qam\}$, $m =
1, 2, ..., n$.
\end{description}
    To each conceivable history (i.e. to any element of a family) one
associates a precise weight which is assumed (if we understand correctly
the aims of
the proponents of the DH approach) to represent, under certain precise
assumptions we will state at short, {\it the
probability that the considered physical system actually possesses, at
the associated times, the properties identified by the projection
operators appearing in it}. The natural candidate for such a probability
is:
\begin{eqnarray}
    p\left[\hisa\right] & = & \tr\{\qan U(\tn, t_{n-1})\qanmu U(t_{n-1},
    t_{n-2})\ldots U(\tu, \tz)W  \nonumber \\
    & & U^{\dagger}(\tu, \tz)\ldots U^{\dagger}(t_{n-1},t_{n-2})\qanmu
    U^{\dagger}(\tn, t_{n-1})\qan\}. \label{2.6}
\end{eqnarray}
Note that this choice amounts to attach to the considered history the
probability that standard quantum mechanics assigns to the process in
which the system under consideration is subjected, at the chosen times,
to the specified measurements, and one always gets the outcome $+1$ for
the associated projectors. In this way one ensures from the very
beginning that the theory is predictively equivalent to standard quantum
mechanics with the orthodox interpretation.
In order that the proposed probabilistic interpretation be tenable,
$p\left[\hisa\right]$ must satisfy the usual probability rules; this happens
iff   Griffiths' condition \cite{ref4}:
\begin{eqnarray}
     &  & \re[\tr\{\pan U(\tn, t_{n-1})\panmu U(t_{n-1},
    t_{n-2})\ldots U(\tu, \tz)W  \nonumber \\
    & & U^{\dagger}(\tu, \tz)\ldots U^{\dagger}(t_{n-1},t_{n-2})\ptanmu
    U^{\dagger}(\tn, t_{n-1})\ptan\}] = 0 \label{2.8}
\end{eqnarray}
holds whenever at least one of the elements of the sequence
$(k, j,\ldots)$
differs from the corresponding element of the sequence
$({\tilde{k}} ,{\tilde{j}},\ldots)$.
Condition (\ref{2.8}) is known as the {\it consistency condition}
\cite{ref4} or the {\it weak decoherence condition}
\cite{ref12}. If it is satisfied, the
corresponding family of histories $\fam$ is said to be consistent (or
weakly decoherent).

    In ref. \cite{ref11}, Gell--Mann and Hartle have introduced a stronger
consistency condition, sometimes called the {\it medium decoherence
condition}:
\begin{eqnarray}
    D(k,j,...; {\tilde{k}},{\tilde{j}},...) &  = &
    \tr\{\pan U(\tn, t_{n-1})\panmu U(t_{n-1},
    t_{n-2})\ldots U(\tu, \tz)W  \nonumber \\
    & & U^{\dagger}(\tu, \tz)\ldots U^{\dagger}(t_{n-1},t_{n-2})\ptanmu
    U^{\dagger}(\tn, t_{n-1})\ptan\} = \label{2.9} \\
    & = & \delta_{k, {\tilde{k}}}\delta_{j, {\tilde{j}}}...
    D(k,j,...;k,j,...), \nonumber
\end{eqnarray}
with obvious meaning of the symbols. The quantity
$D(k,j,...; {\tilde{k}},{\tilde{j}},...)$ itself is
called the {\it decoherence functional}. We point out that while the medium
decoherence condition requires that the off--diagonal elements of the
decoherence functional vanish, the weak decoherence condition
requires only that their real parts vanish\footnote{Actually, Gell-Mann
and Hartle have considered \cite{ref12} also a stronger decoherence
condition.}.
From now on we will always
use condition (\ref{2.9}) and we will refer to it, for simplicity, as the
{\it decoherence condition} without any further specification. Accordingly,
we introduce the following
definition: {\it a family of histories is said to be decoherent if its
fine--grained histories satisfy condition} (\ref{2.9}).

    It is useful to remark that the decoherence condition is a quite
strict requirement on $\fam$ and that in almost all cases (even many of
the physically interesting ones) a family of histories is never exactly
decoherent. For this reason Gell--Mann and Hartle themselves [12]
introduced the idea of approximate decoherence, amounting to require
that:
\begin{equation}
    D(k,j,...; {\tilde{k}},{\tilde{j}},...) \approx
    \delta_{k, {\tilde{k}}}\delta_{j, {\tilde{j}}}...
    D(k,j,...;k,j,...). \label{2.10}
\end{equation}
    In this paper we will analyse only the case of exactly decoherent
histories; the analysis of the case of approximate decoherence and its
physical consequences will be the subject of a subsequent paper.

\subsection{The physical meaning of the decoherence condition.}

    The requirements embodied in condition (\ref{2.8}) or (\ref{2.9}), i.e.
 that quantum probabilities behave like classical probabilities,
are necessary if one takes into account that the interpretation
pretends to make probabilistic claims about properties which are
exhaustive, mutually exclusive and which are claimed to be
possessed by a physical system independently of any observer, or
any measurement process. However, as natural as they can appear,
such requests are not satisfied, in general, by the histories
constituting a family. That this is the case can be easily
understood by taking into account the peculiar probabilistic
structure of standard quantum mechanics. One can exhibit a quite
elementary example which makes this point perfectly clear by
making reference to the famous two--slits experiment, {\it the
only mystery of quantum mechanics}, in Feynman's words
\cite{ref19}.

    Suppose one has a particle in an initial state corresponding to a
wave function which is appreciably different from zero and almost
constant on a large interval in the $x$--direction orthogonal to the
direction along which the particle propagates and which impinges on a
screen with two slits $\delta_{1}$ and $\delta_{2}$ of equal extension
around the points $x_{1}$ and $x_{2}$.
Let us denote by $C$ the complement in the real axis $x$ of the set
$\delta_{1}\cup\delta_{2}$.
We also consider an infinite sequence $\{\Delta_{j}\}$ of disjoint
intervals covering the whole $x$--axis. Let us denote by $P_{\delta_{1}}$,
$P_{\delta_{2}}$, $P_{C}$ and $P_{{\Delta}_{j}}$
the operators projecting on the closed linear manifolds of the
square--integrable functions of $x$ with support entirely contained in the
indicated intervals. Finally we denote as $P_{\delta_{1}\cup\delta_{2}}$
the coarse--grained
projection operator on the manifold associated to the set
$\delta_{1}\cup\delta_{2}$.
Let us now consider the family $\fam^{\tiny\makebox{(2slits, 2times)}}$
of two--times histories characterised by  $\{\pau\} \equiv (P_{\delta_{1}},
P_{\delta_{2}}, P_{C})$
and $\{P^{(2)}_{j}\} \equiv \{P_{{\Delta}_{j}}\}$. Such a
family contains the coarse-grained history:
\begin{equation} \label{2.11}
    \His^{([\delta_{1}\cup\delta_{2}]\&{\Delta}_{j})} =
    \{(P_{\delta_{1}\cup\delta_{2}},\tu), (P_{{\Delta}_{j}},\td)\}
\end{equation}
as well as the fine grained histories:
\begin{equation} \label{2.12}
    \His^{(\delta_{1}\&{\Delta}_{j})} = \{(P_{\delta_{1}},\tu),
    (P_{{\Delta}_{j}},\td)\},\qquad
    \His^{(\delta_{2}\&{\Delta}_{j})} = \{(P_{\delta_{2}},\tu),
    (P_{{\Delta}_{j}},\td)\}.
\end{equation}
It is then obvious that, due to quantum interference:
\begin{equation} \label{2.13}
    p(\His^{([\delta_{1}\cup\delta_{2}]\&{\Delta}_{j})}) \quad\neq\quad
    p(\His^{(\delta_{1}\&{\Delta}_{j})}) +
    p(\His^{(\delta_{2}\&{\Delta}_{j})}),
\end{equation}
i.e. the considered family $\fam^{\tiny\makebox{(2slits, 2times)}}$
is not decoherent.

    Note that the relation (\ref{2.13}), implying that the decoherence
condition is not satisfied, expresses the well known fact that the
probability of a particle being within a certain interval at time  $\td$
subsequent to its passage through the screen with two slits, does not
coincide with the sum of the probability of being within the considered
interval having been within the slit $\delta_{1}$  at time $\tu$,
plus the probability of being within the considered interval having been
within the slit $\delta_{2}$  at the same time.

    Before concluding, we remark that, contrary to what happens in
the case just considered, the families: $\fam^{\tiny\makebox{(2slits,$\tu$)}}$
consisting of one-time histories at time $\tu$ whose
 fine grained elements are identified by the set $\{\pau\} \equiv
\{P_{\delta_{1}}, P_{\delta_{2}}, P_{C}\}$,
as well as the family $\fam^{\tiny\makebox{(2slits;$\tu,\td$)}}$
consisting of
histories at times $\tu$  and $\td$ whose  fine grained elements are
identified by the sets $\{\pau\} \equiv \{P_{\delta_{1}\cup\delta_{2}},
P_{C}\}$
and $\{P^{(2)}_{j}\} \equiv \{P_{{\Delta}_{j}}\}$,
respectively, are decoherent.

    As the reader has certainly grasped, the DH approach allows to
claim, by making reference to  $\fam^{\tiny\makebox{(2slits,$\tu$)}}$,
that the particle either goes
through slit 1 or through slit 2 or it is absorbed. Similarly, by making
reference to $\fam^{\tiny\makebox{(2slits;$\tu,\td$)}}$,
one can state that the particle is either
absorbed at time $\tu$ or it is, at time  $\td$, within one of the intervals
$\Delta_{j}$. But the fact that $\fam^{\tiny\makebox{(2slits, 2times)}}$
is not a decoherent family, does not
allow to make statements of the kind: the particle is now within the
interval $\Delta_{j}$ and it went through slit 1 (slit 2). We stress that
the just
outlined distinction between $\fam^{\tiny\makebox{(2slits,$\tu$)}}$ and
$\fam^{\tiny\makebox{(2slits;$\tu,\td$)}}$ is an instance of the `single family
rule', i.e. of the fact that quantum reasoning must always deal with a single
decoherent family.

\section{The logic of decoherent families.}

In this Section we discuss the way in which some of the proponents of the
DH approach
have suggested to give a precise and (as far as possible)  classical
logical structure
to the theory. This is done by equipping the histories of any decoherent
family with
a Boolean structure and by defining (taking into account the probabilistic
nature of
the scheme) the concept of logical implication between histories of a
decoherent
family.

\subsection{Decoherent families and Boolean algebras.}

    Omn\`es \cite{ref8}, and subsequently Griffiths \cite{ref6,ref7}, have
shown that any family of histories is naturally
equipped with a Boolean algebraic
structure. The extreme relevance of such a step for the DH
approach derives from the fact that it is a condition which should allow to
recover classical logic and classical reasoning within a quantum
context, avoiding in this way the problems arising in connection with
quantum logic. Obviously, even though one
can, formally, give an algebraic Boolean structure to any family, only in
the case of decoherent families the probabilities defined by Eq. (\ref{2.6})
define a classical probability measure on the set of histories, and, as
such, they allow to give a physical meaning to the histories themselves.

    Here we have no need to discuss this
problem in its full generality; for our purposes it will be sufficient to
consider a family  characterized by a unique time instant  $t$:
\begin{equation}
\His^{(\alpha)} = \{(Q_{\alpha}, t)\},
\end{equation}
where $\{Q_{\alpha}\}$ is the set of projection
operators specifying the histories of the family. In this simplified case
the conjunction and disjunction of two histories, and the negation of a
given history,
 i.e. the logical connectives necessary to build a Boolean algebra, are
defined according to:
\begin{eqnarray}
\His^{(\alpha)}\wedge\His^{(\beta)}  =
\{(Q_{\alpha}\wedge Q_{\beta}, t)\}
& \qquad & Q_{\alpha}\wedge Q_{\beta} =
Q_{\alpha}Q_{\beta}, \nonumber \\
\His^{(\alpha)}\vee\His^{(\beta)}  =
\{(Q_{\alpha}\vee Q_{\beta}, t)\}
& \qquad & Q_{\alpha}\vee Q_{\beta} =
Q_{\alpha} + Q_{\beta} -
Q_{\alpha}Q_{\beta},
\\
\His^{(\alpha)^{\perp}}\quad =\quad\,\,\, \{(Q_{\alpha}^{\perp},
t)\} & &
Q_{\alpha}^{\perp} = 1 - Q_{\alpha}. \nonumber
\end{eqnarray}
It is an elementary task to show that all conditions characterizing a
Boolean algebra are satisfied.

\subsection{The logical implication between histories.}

As already stated, to work out a consistent `quantum reasoning' referring to
histories one has to introduce in a mathematically precise way the idea of
one history
of a decoherent family implying (a fact we will denote by
the symbol ``$\Longrightarrow$'') another history of the same family. This
is easily
done by following the standard way used within a probabilistic context:
\begin{equation}
\hisa\,\Longrightarrow\,\hisb\quad\makebox{if and only if:}\quad
\frac{p[\hisa\,\,\wedge\,\,\hisb]}{p[\hisa]} = 1.
\end{equation}

For future purposes, we call the attention of the reader on the fact that
the above
definition makes exclusive reference to the probabilities that the formalism
attaches to the decoherent histories.

\subsection{An illuminating discussion.}

When one adopts the above position about logical implication one is immediately
confronted with a problem which can be summarized as follows: the same
history, when
considered as belonging to different decoherent families, can lead to
contradictory
implications. In fact, it is easy to show that one can identify two compatible
histories\footnote{i.e. such that
there is a decoherent family  to which they both belong.}
$\hisa$ and $\hisb$ belonging to a certain decoherent family $\fam_{1}$,
such that $\hisa$  implies  $\hisb$.
Furthermore one can also identify   a second decoherent family $\fam_{2}
\neq \fam_{1}$
such that the original history $\hisa$  and another history $\hisg$
belong to it and, within $\fam_{2}$,  one can prove that $\hisa$ implies
$\hisg$. We then have:
\begin{eqnarray}
    & & \hisa,\hisb\,\in\,\fam_{1};\qquad \hisa\,\Longrightarrow\,\hisb
    \label{5.1a} \\
    & & \hisa,\hisg\,\in\,\fam_{2};\qquad \hisa\,\Longrightarrow\,\hisg
    \label{5.1b}
\end{eqnarray}
Given the above facts one could naturally be tempted to conclude that $\hisa$
implies {\it both
$\hisb$ and $\hisg$}:
\begin{equation} \label{2.inc}
    \hisa\,\Longrightarrow\,\left[\hisb \wedge \hisg\right],
\end{equation}
proving in this way the inconsistency of the theory, since
one can easily devise histories satisfying the above
conditions but such that $\hisb$  and $\hisg$ make reference to physical
properties
 which  cannot be simultaneously true.
Formally, $\hisb$ is associated to a certain projection operator $B$ while
$\hisg$ is associated to another such operator $C$, where $BC  = CB = 0$, but
 $B \neq 1 - C$. Examples of this type have been considered for the
first time by  Aharonov and Vaidman \cite{AV} and subsequently by
Griffiths and Hartle \cite{gh} in connection with Kent's criticism
\cite{ref18, ken} of the DH-approach.

A way to circumvent the just outlined difficulty has been suggested by
Griffiths himself who, in his most recent papers \cite{ref6, ref7}, has
repeatedly
stressed the fundamental relevance of the {\bf ``single family rule''}
(or logic, or framework)  we have mentioned previously \cite{ref7}:
\begin{quotation}
    {\small From now on we shall adopt the following as the fundamental
    principle of quantum reasoning: \it A meaningful description of a
    (closed) quantum--mechanical system, including its time development,
    must employ a single framework}.
\end{quotation}

It seems to us that the most obvious way in which one has to
understand the above rule is that when one resorts to logical
tools to draw conclusions about histories he has to deal only with
histories belonging to the same decoherent family. If one agrees
on this point, to state that  ``$\hisa$ implies {\it both $\hisb$
and $\hisg$}'' is incorrect, since $\hisb$ and $\hisg$  belong to
incompatible families. Stated differently, since there is no
decoherent family which can accommodate both of them,  there is no
consistent way of attaching a meaning to the conjunction of the
two histories.

    In our opinion, the fact that, while one can claim that
$\hisb$ and $\hisg$  are  {\it separately} true
(given for granted the truth of $\hisa$) it
is nevertheless meaningless to consider them {\it together},
is rather puzzling. How can one accept this peculiar aspect
of the formalism? The situation becomes even more embarrassing
when one takes into account that the two histories
$\hisb$ and $\hisg$ may very well (as we have already said)
make claims about properties
which are mutually incompatible. Summarizing, we are inclined to agree with
d'Espagnat
when he claims that the situation we have just analyzed
represents a real logical paradox \cite{ref15}. In spite of the above remarks
we are perfectly aware that, at the {\it purely formal}
level, one can claim that the DH approach, when enriched by the single
family rule
interpreted in the above sense, is free from the just mentioned
inconsistencies.

\section{How many decoherent families are possible?}

    As mentioned in the introduction, there is a serious source
of difficulties in the theory, coming from the fact that there are many
decoherent families --- actually, most of them --- which are manifestly
devoid of any physical meaning. For example, it is easy to identify (and
Griffiths himself has presented \cite{ref6} an explicit example of such a
situation) decoherent families in which Schr\"{o}dinger's cat is not {\it
dead or
alive}, but {\it dead + alive}.
Actually, Giardina and Rimini \cite{ref21} have proved
that starting from a quasi--classical decoherent family it is possible to
build infinitely many inequivalent decoherent families which are, in
general, highly non--classical.
A question then naturally arises: which physical interpretation can be
given to this type of decoherent families?

    The proponents and the supporters of the decoherent
histories approach have already identified this problem and have taken
various positions about it. For example {\bf Griffiths} has repeatedly  stated
\cite{ref7} that {\it all decoherent families have to be put on the
same footing}. In his words:
\begin{quotation}
{\small The formalism allows the physicist to choose from a very
large number of alternative frameworks or consistent families of
histories, all of which are considered ``equally valid'' in the sense that
no fundamental physical law determines which family should be used in
any given case.}
\end{quotation}
 Accordingly, facing the problem of the existence of
what we have characterised as {\it physically senseless decoherent families}
he stated \cite{ref6}:
\begin{quotation}
{\small What would happen if, ten minutes from now, we were to abandon the
quasi--classical framework for one in which, say, there is a coherent quantum
superposition of the computer in distinct macroscopic states? Of course,
nothing
particular would happen to anything inside the box; we, on the other hand,
would
no longer be able to describe the object in the box as a computer, because the
language consistent with such a description would be incompatible with the
framework we were using for our discussion.}
\end{quotation}
In previous papers \cite{ref4}, he had also stated:
\begin{quotation}
{\small Such histories [i.e. the senseless ones]
do exist, and the consistent histories
approach will assign probabilities to consistent families of {\bf grotesque}
histories, if that is what interests the theoretician. The point we wish to
make is not that {\bf grotesque}  events are somehow ruled out by the
consistent histories approach (obviously they are not), but simply that
they are not an essential part of interpreting what happens in an
{\bf ordinary}  consistent history.}
\end{quotation}
    In our opinion Griffiths' reasoning is shifty and weak: what is an
{\it\bf ordinary} family? What distinguishes it from the other families? Why
should a theoretician consider only ordinary families? Alternatively, if
he wants to consider also non-ordinary families, which physical
meaning will he attach to them? Such fundamental questions are left
unanswered.

    {\bf Omn\`es} \cite{ref9} has made a serious attempt to give a precise
meaning
to the idea of a sensible decoherent family stating explicitly:
\begin{quotation}
{\small One must
first restrict oneself to a special class of logics [i.e. decoherent
families]:  those containing all the actual facts, i.e. all the real classical
phenomena ... they may be said to be sensible.

One can then assert
what should be true. To begin with, actual facts will be taken to be true.
Some other properties ... will also said to be true when they satisfy the
two following criteria: \\
--- One can add them to {\it any} sensible logic while preserving consistency.
\\
--- In all these augmented logics, (they are) logically equivalent to a
factual phenomenon.}
\end{quotation}

    In this way he pretends to prove that only decoherent histories
referring to definite macroscopic properties (past, present and future),
as well as all histories referring to the properties of microsystems
subjected to measurement processes are true. Thus, Omn\`es
does not commit himself about the truth or falsity of
the histories of all decoherent families, but only of those histories which are
related to ¦``classical''² phenomena in accordance with the above criteria.
However,
Dowker and Kent \cite{ref14} have shown that his argument is not satisfactory.

   {\bf Gell--Mann and Hartle} \cite{ref11} proposed to consider an appropriate
{\it measure of classicity}  as a basic criterion to pick out the most
classical
family (or families) among all the decoherent ones:
\begin{quotation}
{\small The impression that there is something like a classical
domain suggests that we try to define quasiclassical domains
precisely by searching for a measure of classicity for each of the
maximal sets of alternative decohering histories and concentrating
on the one (or ones) with maximal classicity.}
\end{quotation}
Unfortunately, as they admitted \cite{ref12}:
\begin{quotation}
{\small ... we have not so far
progressed beyond as to how that could be done.}
\end{quotation}
Moreover, they have
not made sufficiently clear the role which has to be given to the
quasi-classical
 families which differ from the ones we experience. Natural
questions arise: have such families to be considered as meaningless or
do they correspond to other aspects of reality to which we
have no access? It seems that
these authors prefer this second alternative \cite{ref25}:
\begin{quotation}
{\small Among all the possible sets of alternative histories for
which probabilities are predicted by the quantum mechanics of the
universe, those describing a quasiclassical realm [i.e. quasiclassical
decoherent families]  like the one that includes familiar experience, are
of special importance ... Such quasiclassical realms are important for
at least two reasons ... [The second one being that]  coarse--grainings of
this usual quasiclassical realm are what we (humans and many other
systems) use in the process of gathering information about the universe
and making predictions about its future.}
\end{quotation}

    In our opinion one should put forward reasons to explain why
we human beings do not perceive the existence of other domains. Gell--Mann
and Hartle propose \cite{ref11} the following explanation:
\begin{quotation}
{\small If there are many
essentially inequivalent quasiclassical domains, then we could adopt a
subjective point of view, as in some traditional discussions of quantum
mechanics, and say that the IGUS\footnote{An IGUS --- Information Gathering and
Utilizing System --- is a
complex adaptive system (as, for example, a human being) able to interact with
the surrounding environment, gather and elaborate information coming from it.}
``chooses'' its coarse--graining of
histories and, therefore, ``chooses'' a particular quasiclassical domain,
or a subset of such domains, or further coarse grainings. It would be
better, however, to say that the IGUS evolves to exploit a particular
quasiclassical domain or set of such domains.}
\end{quotation}
In view of the fact that the possibility of limiting the set of allowed
families has already been entertained even by some of the supporters of the DH
approach, it will be quite natural to take the same position to overcome the
problems we are going to focus in the second part of this paper.

\vspace{1cm}
\begin{center}
{\Large\bf  Part II}\\
\end{center}
\vspace{0.5cm}

As already anticipated, this part will be devoted to a critical investigation of
the nature of the probabilistic statements of Decoherent Histories. A nice way
to stress this crucial point  derives from taking into
consideration the sharp remarks by J. S. Bell \cite{ref2} about the
ortodox interpretation of Quantum Mechanics:
\begin{quotation}
    {\small In the beginning, Schr\"{o}dinger tried to interpret his
    wavefunction as giving somehow the density of stuff of which the world
    is made. He tried to think of an electron as represented by a wavepacket
    ... a wevefunction appreciably different from zero only over a small
    region in space. The extension of that region he thought of as the
    actual size of the electron ... his electron was a bit fuzzy. At first
    he thought that small wavepackets, evolving according to the
    Schr\"{o}dinger equation, would remain small. But he was wrong.
    Wavepackets diffuse, and with the passage of time become indefinitely
    extended, according to the Schr\"{o}dinger equation. But however far the
    wavefunction has extended, the reaction of a detector to an electron
    remains spotty. So Schr\"{o}dinger's ``realistic'' interpretation of his
    wavefunction did not survive.

    Then came the Born interpretation. The wavefunction gives not the
    density of {\it stuff}, but gives rather (on squaring its modulus) the
    density of {\it probability}. Probability of what, exactly? Not of the
    electron {\it being} there, but of the electron being {\it found} there,
    if its position is ``measured''.

    Why this aversion to ``being'' and insistence on ``finding''? The
    founding fathers were unable to form a clear picture of things on the
    remote atomic scale. They became very aware of the intervening
    apparatus, and of the need for a ``classical'' base from which to
    intervene on the quantum system. And so the shifty split.}
\end{quotation}
The proponents of the DH approach,
in their first papers, seemed to share the
worries so nicely expressed by Bell and to be mainly motivated by the desire
to avoid giving any privileged status to measurement processes and observers.
It is therefore interesting to discuss whether these requirements can be 
embodied within
the probabilistic framework of Decoherent Histories.  We begin by observing that 
Bell's
pretension that the probabilities of a theory refer, e.g., to an electron 
``being'' in
a given place, is logically equivalent to requiring that statements like ``the
electron is confined within a given region'' be either true or false
independently of any further specification and of the fact that we only know
how probable is that they are true.

    We will now try to analyze in great detail the problem of the truth
values of the probabilistic statements of the theory.

\section{Probabilities and truth--values.}

    As we have already said, the basic assumptions
of the DH approach is that,
given a family of histories satisfying the decoherence condition,
the diagonal elements of the decoherence functional define a probability
distibution on the histories of the considered family:

\begin{quotation}
{\small Provided a consistency condition is satisfied,
the corresponding Boolean algebra of events, called a framework,
can be assigned {\it probabilities} in the
usual way, and within a single framework quantum reasoning is identical to
ordinary probabilistic reasoning. (Griffiths, \cite{ref6})}
\end{quotation}
\begin{quotation}
    {\small {\it Probabilities} (approximately obeying the rules
    of probability theory) can
    be assigned only to sets of histories that approximately decohere.
    (Gell--Mann and Hartle, \cite{ref11})}
\end{quotation}
But then the question put forward by John Bell arises:
{\it probability of what, exactly?}
Before giving an answer to this apparently elementary question let us
comment in general on the meaning of probabilities in Classical and
in Quantum Mechanics, by discussing the two following examples:
\begin{enumerate}
    \item[a)] Let us consider the tossing of a coin; the sample space has
    two
    elements, i.e. the two significative events
    ``head'' $H$ and ``tail'' $T$. The probabilities
    assigned to these events are defined in accordance with the assumption
    that
    the coin is fair:
\begin{equation}
    p(H) = 1/2, \qquad p(T) = 1/2.
\end{equation}
These assignements define all statistical properties concerning the tossing
of the coin.

\item[b)] Let us now take into account the spin properties referring to the
component along the $z$-axis of a spin  $1/2$ particle in the state
$|x\uparrow\rangle$. Also
in this case the sample space has two elements, which are
the two physically meaningful events  ``spin up'' $z\uparrow$ and
 ``spin down'' $z\downarrow$. The associated probabilities, in accordance
with the quantum mechanical rules, are:
\begin{equation}
    p(z\uparrow) = 1/2, \qquad p(z\downarrow) = 1/2.
\end{equation}
Also in this case, the considered assignment characterizes all statistical
properties concerning the spin of the
particle along the $z$-axis.
\end{enumerate}
It is obvious that, from the purely probabilistic points of view,
the two cases are perfectly equivalent. {\it What is then the
profound difference between Classical and Quantum Mechanics?} It
resides uniquely in the radically different roles the two
formalisms attach to probabilities. According to Classical
Mechanics, the probabilities refer to objective properties (i.e.,
properties which are intrinsic to the system, assigned {\it
a-priori} independently from any act of observation) of the
physical systems under consideration, while Quantum probabilities
refer exclusively to the possible outcomes of appropriate
measurement procedures conditional under the measurements being
performed. This means that physical properties do not exist by
themselves, independently from measurement procedures. All this is
well known, but the relevant question is: {\it is it possible to
give a mathematically precise formal expression to this radical
difference between the Classical and the Quantum cases?} The
answer is also well known: the appropriate formal way to
characterize the difference derives from the assignment of truth
values to the various events. When we deal with classical
probabilities, i.e. with probabilities referring to intrinsic,
objective properties of physical systems, we can assign to each
element of the sample space, and in general to each event of the
corresponding Boolean algebra, a truth value:
\begin{eqnarray}
& & \makebox{In Classical Mechanics:} \nonumber \\ & & \exists\,\,
h: \{x\,\in\,\makebox{Boolean Algebra of events}\} \longrightarrow
\{0,1\}.
\end{eqnarray}
Obviously, in general we do not know which truth value has to be
attached to any event since, in the general case, we have only an
incomplete knowledge of the actual physical situation. This fact,
however, is totally irrelevant from the ontological point of view:
if we believe that Classical probabilities refer to properties
objectively possessed by physical systems, properties which exist
independently from us and our knowledge, then we can legitimately
state that any event (the mathematical counterpart of the physical
properties) is associated to a precise truth value: it is
``true''(1) when it corresponds to properties actually possessed
by the physical system, it is ``false''(0) if it describes
properties which are not possessed by the physical system.
Moreover, we can  identify the essential features that the
correspondence $h$ must exhibit. Technically $h$ must be a
two--valued homomorphism from the Boolean Algebra formed by the
subsets of the sample space onto the Boolean Algebra $\{0,1\}$.
This means that the application $h$ must preserve the join, meet
and complement relations:
\begin{eqnarray}
    & i. & h(x \vee y) = h(x)\vee h(y), \label{4.2a} \\
    & ii. & h(x \wedge y) = h(x)\wedge h(y), \label{4.2b} \\
    & iii. & h(x^{\perp}) = h(x)^{\perp}, \label{4.2c}
\end{eqnarray}
for every $x, y$ belonging to the Boolean Algebra.
Obviously, such requests must be satisfied if we want to be allowed to
use classical reasonings about classical events.
For example, suppose $x$  is a true event:
$h(x) = 1$. Then, the classical way of reasoning requires its negation
to be false: $h(x^{\perp}) = 0 = 1^{\perp} = h(x)^{\perp}$.
Similarly, if $x$  and $y$
are two events which are true and false, respectively --- $h(x) = 1$,
$h(y) = 0$  --- then their
join must be true and their meet must be false:
\begin{eqnarray}
    & & h(x \vee y) = 1 = 1\vee 0 = h(x)\vee h(y), \label{4.3a} \\
    & & h(x \wedge y) = 0 = 1\wedge 0 = h(x)\wedge h(y), \label{4.3b}
\end{eqnarray}
and similarly for all other cases.

As we have stated, within Quantum Mechanics, systems do not
possess objective properties and probabilities refer simply to
measurement outcomes. In such a context, one cannot, in general,
attribute {\it any} truth value to the events which are correlated
with the  ``properties'' (i.e. with the projection operators) of
quantum systems:
\begin{eqnarray}
& & \makebox{In Quantum Mechanics:} \nonumber \\ & &
\not\exists\,\, h: \{x\,\in\,\makebox{Boolean Algebra of events}\}
\longrightarrow \{0,1\}.
\end{eqnarray}

After these remarks, we can come back to our analysis of
Decoherent Histories; in this context, the elements of the sample
space, and the events of the Boolean algebra associated to it, are
the histories belonging to a single family, and, when the
decoherence condition is satisfied,  a probability distribution on
such histories is defined. In this case too, when one limits his
considerations to a single family of decoherent histories, the
situation is, from a formal point of view, identical to the one of
the Classical and of the Quantum cases.
 However,
to have a complete picture one must give a precise answer to the original
question:
{\it the probabilities which the DH approach attaches to any decoherent history
(which we will denote as DH-probabilities) what do they refer to?} To
properties
objectively possessed by physical systems or to measurement outcomes, just
as the Quantum
probabilities? Obviously, since the purpose of the DH--approach is that of
solving the interpretative difficulties of Quantum Mechanics and to put
forward a realistic theory, the
DH-probabilities, just like Classical ones, must refer to objective and
intrinsic properties of physical systems: there is no other reasonable
alternative.
But, if this is the situation, if these are the aims of the theory,
our previous analysis leads to conclude that any decoherent history must
have a precise
truth value, independently from the fact that, in general, we have only a
probabilistic knowledge about the system. Moreover, the relation
between histories and truth values must be a homomorphism:
\begin{eqnarray}
& & \makebox{DH--probabilities like Classical probabilities
$\Rightarrow$} \nonumber \\
& & \exists\,\, h: \{\His\,\in\,\makebox{Decoherent $\fam$}\}
\longrightarrow \{0,1\}, \\
& & \makebox{$h$ is a homomorphism.} \nonumber
\end{eqnarray}

    In the second section of this paper we have mentioned that, within the
DH-approach any reasoning, any logical argument, must be entirely developed
within the context of a single family of decoherent histories. If this rule
is not
respected one risks to draw contradictory conclusions.
At first sight, this feature of the theory {\it might} be considered as
implying
 that every decoherent family has its precise homomorphism, independently
from the truth
values attached to histories belonging to decoherent families which are
incompatible
with the previous one. If this is the situation, it {\it would} be more
appropriate to
denote the homomorphism as $h_{\tiny\fam}(\His)$.
This notation  makes evident
the fact that classical reasoning is valid {\bf iff} it is confined to a
single
decoherent family, and, as such, it depends crucially on the family we
are dealing with. To better grasp this important point one  could make
reference, once more, to the histories about the two--slits experiment
discussed in Section 2. In such a case with reference, e.g., to
$\fam^{\tiny\makebox{(2slits,$\tu$)}}$,
one can state that the particle goes {\it either} through slit 1 {\it or}
through slit 2 and that if it is true that it goes through slit 1 then it is
false that it goes through slit 2 and viceversa. Similarly, the
probabilistic structure of the set of histories of
$\fam^{\tiny\makebox{(2slits;$\tu,\td$)}}$ allows one
to claim that a particle which has not been absorbed by the screen, is
objectively within one and only one of the disjoint intervals $\Delta_{j}$,
i.e.,
only one of these histories turns out to be true, all the remaining ones
being false. In spite of this, as repeatedly stressed, statements asserting
through which slit the particle went and which final position it has are
not legitimate since such histories belong to the family
$\fam^{\tiny\makebox{(2slits, 2times)}}$
which is not decoherent and as such no truth value can be attached to
them.

To conclude this Section, we consider it appropriate to analyze
some statements by Griffiths about the truth values which one can
attach to decoherent histories \cite{ref7}:
\begin{quote}
    {\small One important difference between Omn\`es and CHQR [Consistent
    Histories and Quantum Reasoning] is in the definition of ``true''. In
    CHQR, ``true'' is interpreted as ``probability one''. Thus if certain
    data are assumed to be true, and the probability, conditioned upon
these
    data, of a certain proposition is one, then this proposition is true.
    The advantage of this approach is that as long as one sticks to a
single
    framework, ``true'' functions in essentially the same way as in
ordinary
    logic and probability theory. However, because probabilities can
only be
    discussed within some framework, comparison of ``true'' between
    incompatible frameworks are impossible, and in this sense ``true''
    interpreted as ``probability one'' must be understood as relative to a
    framework. \\
    The feature just mentioned has been criticized by d'Espagnat. But it is
    hard to see how to get around it if one wishes to maintain (as do
Omn\`es
    and I) that reasoning inside a single framework should follow classical
    rules, and classical rules associate ``true'' (in a probabilistic
    theory) with ``probability one''.}
\end{quote}
Such statements seems to us to confuse the reader rather than clarifying
the matter. In fact:
\begin{itemize}
    \item In the usual physical language the statement that ``something is 
true''
makes reference to {\it objective} properties of a physical system, to some
{\it elements of physical reality}, in Einstein's language. If Griffiths
resorts to the expression  ``true'' having in mind something different he
should be very precise about it, or even resort to  a different term.

\item In  Classical Statistical Mechanics, while
``true'' is  correctly associated with ``probability one'', it is not the case
that ``true'' is associated {\it only} with ``probability one''. Within such a
theory {\it any} proposition has a precise truth value quite
{\it independently} from the probability which is attached to
it\footnote{Exception made for the fact that, if its
probability equals one, then the proposition must  obviously be true, while if
it equals zero, then it must be false.}. As a consequence, when
Griffiths writes: {\it in CHQR, ``true'' is interpreted as ``probability
one''}, he must choose one and only one of the two following alternatives:
\end{itemize}
\begin{enumerate}
    \item Just as in the case of Classical Statistical mechanics, if the
theory associates to a proposition probability one (zero) then the
proposition is certainly true (false), while in all other
instances {\it the proposition   has in any case a truth value,
which, however, is unknown to us}. If this is the case then {\it
every decoherent history has a truth value} (which, as previously
discussed can be formally described by an appropriate
homomorphism). We will analyse in what follows the implications of
taking such a position.
    \item {\it Only the decoherent histories having probability one are
true and those having probability zero are false, while all other
histories do not possess a truth value.} If this is the case, then Griffiths
is discriminating some decoherent histories from the remaining ones and he is
giving a particular conceptual status only to some of them.
Some histories have a
precise physical meaning and ontological status, since they have a truth
value, while all the remaining ones have no physical meaning whatsoever. Such
a position is perfectly legitimate; however we cannot avoid remarking that
 ``in the real world'' probabilities which are exactly equal to one never
occur\footnote{Leaving aside some pathological and physically
uninteresting cases.}. Accordingly,
there is always a precise (possibly very small) probability that
different things might happen. For instance, the wavefunctions of all
physical systems have non compact support in configuration space (exception
made for single time instants). Consequently, a physical system which is at a
certain place at a given instant can be, at any subsequent time, at an
arbitrarily far away position (even though the probability of such an event is
extremely small in the case of macroscopic objects). In simpler terms, the
fact that a table is here now does not imply, with probability equal to one,
that it will still be here within a second. The conclusion seem to us
unavoidable: if this is what Griffiths has in mind, then   {\it nothing is
true and nothing is false}, since nothing has probability one or zero
of occurring.

Moreover, within Standard Quantum Mechanics itself it is perfectly legitimate
and consistent to attach a definite truth value to events (or equivalently to
histories) which have probability one or zero to actually occur. There is no
need to resort to a new interpretation, such as the one characterizing
Decoherent Histories, to get this result.

\end{enumerate}

\section{The ``single family'' rule and truth--values.}

In the previous sections of this work we have mentioned that a
basic assumption of the DH approach of Griffiths is the strict request
that any argument must be developed within a {\it single family}
of decoherent histories. If this rule is not respected one risks to derive
(uncorrectly)
contradictory and/or  inconsistent conclusions.
Griffiths himself, in his most recent papers \cite{ref6, ref7}, has repeatedly
stressed the fundamental relevance of the  ``single family rule''.

In this Section we will make an attempt to understand what he means
exactly by requiring that the ``single family rule'' be an essential part of
the theory, and in which cases it must be applied. To clarify this point, we
begin by quoting a long sentence from a paper
\cite{gri1} published in 1987 by this author:
\begin{quotation}
    {\small A sheet of paper weighing 5 g is torn in two, the two pieces are
    placed in opaque envelopes, and one is mailed to a scientist in London
    and the other to a scientist in Paris. \\
    On the basis of this information, the masses of the individual pieces
    of paper are unknown but strongly correlated in that their sum must be 5
    g. Thus, for example, if the envelope in London is opened and the piece
    of paper is weighed and has mass of 3 g, we can at once conclude that
    the piece in Paris has mass of 2 g ... \\
    Needless to say, the fact that one can immediately infer from a
    measurement in London the weight of a piece of paper  hundreds of
    kilometers away in Paris has nothing to do with some strange ``action at
    a distance'' ... Weighing the piece in London has no {\it physical}
    effect upon the piece of paper in Paris.
    What it does affect is our {\it knowledge}
    about the latter, which is something very different. Of
    course, if we fail to distinguish  between what is there and our
    knowledge of what is there, we may be confused into thinking that
    measurement on one object have ``spooky'' effects on a very distant
    object.}
\end{quotation}
In this passage Griffiths identifies correctly
two different levels of reasoning within the context of Classical Physics:
\begin{itemize}
\item On one side there is the {\bf knowledge we have about a given
    physical system}; such a knowledge
    is {\it subjective} since it depends crucially from the information we
    have and from the aspects of the system we are interested in
    considering.
    Moreover, it has, typically, a  {\it probabilistic} nature, since (in
    practice) it is  impossible to determine with infinite precision all
    variables which characterize the state of the system. In the specific
    example chosen by Griffiths, the knowledge by the observer exhibits also
    some ``non--local'' features,
    since it implies an instantaneous change of
    our knowledge about far away systems.

\item On the other side we have the
    {\bf properties objectively possessed by physical systems}. They have
    an {\it objective} status, since they do not depend on us and on our
    acting (such as  performing a measurement process) on the physical
    system: accordingly, they exist {\it a priori}.
    Moreover, with reference to the example considered by
    Griffiths there are no instantaneous changes of
    properties at--a--distance.
\end{itemize}
Obviously, within standard Quantum Mechanics, the distinction we have made
is meaningless simply because within such
a framework there are no properties which are objective and independent from
the measurement processes devised to ascertain them. On the contrary, the above
distinction is perfectly legitimate within Classical Mechanics, and even
within the DH approach, if it is true that the  DH--probabilities, just as the
classical ones, describe objective properties of physical systems.

Let us come back to the  ``single family'' rule: it represents, as
explicitly pointed out by Griffiths, {\it the fundamental rule of  quantum
reasoning}. But let us raise the question:
what precisely is {\it ``quantum reasoning''?} Griffiths writes \cite{ref6}:
\begin{quotation}
    {\small The type of quantum reasoning we shall focus on this section is
    that in which one starts with some information about a system, known or
    assumed to be true, and from these {\it initial data} tries to reach
    valid {\it conclusions} which will be true if the initial data are
    correct ... \\
    Since quantum mechanics is a stochastic theory, the initial data and the
    final conclusions will in general be expressed in form of probabilities,
    and rules of reasoning are rules for deducing probabilities from
    probabilities ...\\
    Since probabilities in ordinary probability theory always refer to some
    sample space, we must embed quantum probabilities referring to
    properties or the time development of a quantum system in an appropriate
    framework.}
\end{quotation}
In \cite{ref7} he adds:
\begin{quotation}
    {\small Choosing a coarse--graining [of phase--space]
    and choosing a quantum framework are analogous in that in both cases the
    choice is one made by the physicist in terms of the physical problems he
    wishes to discuss ... this choice has no influence on the behavior of
    the
    system being described, although it may very well limit the type of
    description that can be constructed.}
\end{quotation}
These quotations, as well as many others, show that what Griffiths
denotes as ``quantum reasoning'', and then also the ``single
family rule'', make exclusive reference to our knowledge about a
physical system and to the way in which we can take advantage of
such a knowledge to derive new information about it. However, in
our opinion, there is a fundamental point which remains obscure:
{\it does the ``single family rule'' refer exclusively to the
knowledge we have about physical systems and to the way one can
take advantage of such a knowledge to infer further information
about them, or does it refer also to the properties which are
objectively possessed by them?} Since in the very formulation of
the rule no reference is made to the just raised question, let us
analyze the two possible scenarios we have outlined. \\

$\bullet$ Let us suppose that the ``single family rule'' applies not only
to our knowledge about physical systems but also to the properties
characterizing them. Let us then consider a history
$\hisa$ referring to the properties of an elementary particle, such as
``its spin points in a certain direction'' (obviously analogous
considerations can be developed with reference to macroscopic
objects). What can be stated about the truth value of such a history?
\begin{enumerate}
    \item The probability associated to the history (as for all actual 
histories)
is neither identically one, nor identically zero. In accordance with the
previous discussion  Griffiths has not been explicit about
whether
$\hisa$
has a truth value or not. If does not the history represents simply a
statement devoid of any sense. In the case of microsystems this could be
accepted, but, as already remarked, just the same situation can occur for
macroscopic systems. In brief the theory cannot make any interesting
statement. Let us then analyze the other possibility, i.e. that the history
has a truth value and that it is, e.g., true.

    \item Since, in agreement with the position we are analyzing, the
``single family'' rule claims that the properties of physical systems are
unavoidably linked to the decoherent family we use to describe them,
the history $\hisa$ does not have  any truth value {\it per se}, but it
acquires it only with reference to the decoherent family we decide to choose.
Such a situation is unprecedented in classical physical theories
where the properties of physical systems are
    ``objective'' and cannot depend on further specifications.

    \item However, a further crucial fact has to be taken into
account. As remarked in the introduction,  the same history belongs, in
general, to different and incompatible decoherent families. Accordingly, let
us consider now a second decoherent family, which is incompatible with the
previous one, but containing
    $\hisa$. Since we have changed the family, the  ``single family'' rule
    (according to the interpretation we are discussing), forbids us to claim
that such a history is necessarily still true, since such a
conclusion was valid within another family. We are, in a sense, back to square
one, just because we have changed the family. Consequently, the history
$\hisa$,  when considered as a member of the second family, might have a
different truth value, it might be false\footnote{If one accepts that the
truth value cannot change when the family is changed one accepts one of the
assumptions we will consider below and which we will show to clash with the
other physically sensible assumptions we will put forward.}.
The fact that
the truth value of a history and accordingly also the properties of physical
systems may change by changing the decoherent family one decides to take into
account does not represent by itself an unacceptably peculiar feature of the 
formalism?
\end{enumerate}
Accepting that the truth value of a precise history belonging to a
decoherent family (and consequently the assertions about
properties objectively possessed by individual physical systems),
could change according to the decoherent family we choose to
describe them, would represent, within the present formalism, the
exact analogue of accepting the contextual nature of most
properties in hidden variable theories (recall that resorting to
contextuality is the appropriate way for such schemes to
circumvent the Kochen and Specker no--go theorem \cite{ref23}). We
are not worried by the fact that some properties of a physical
system could turn out to be contextual, i.e. non objective, since
they depend upon the overall context.

    However, it has to be stressed that while contextuality can be
accepted within hidden variable theories in which it simply means that
the outcomes of measurements depend, besides the ``state''
characterizing the system, also on the way the measurement is carried
on\footnote{Equivalently, one could claim, with Bell \cite{ref26},
D\"{u}rr, Goldstein and Zangh\'{\i} \cite{ref27}, that in
hidden variable theories, typically in Bohmian mechanics, the appropriate
way out from the necessity of accepting contextuality derives
from claiming that what the theory is
about is simply the positions of all particles of the universe at any time.},
the situation in the present case is radically different. Here the
dependence from the context does not refer to {\it different actual situations}
but to {\it different choices} about what we want to assert about our physical
system. It seems to us that accepting such a form of contextuality within
the DH approach would spoil the theory not only of its original
meaning, but actually of any meaning at all. In fact within DH, and contrary
to the case of hidden variable theories, there is no
way to divide properties into noncontextual and contextual, so that, if one
gives up the assumption that the truth value of a given history does not
depend from the perspective we choose to speak about the system, one must
consider all properties as contextual. On the other hand, as D\"{u}rr,
Goldstein and Zangh\'{\i}
\cite{ref27} have appropriately stressed:
\begin{quotation}
    {\small  Properties that are merely contextual
    are not properties at all; they do not exist, and their
    failure to do so is in the strongest sense possible.}
\end{quotation}

$\bullet$ Let us suppose now that the ``single family rule''
refers only to our knowledge of physical systems and to the way we
can make use of such a knowledge to get new information about
them. In this case, it is  obvious why ``quantum reasoning'' turns
out to have, in general, a probabilistic nature and why it is
subjective, (in the sense that it depends in a crucial manner from
the information we have and from what we are interested in
knowing), but these facts do not, in any way and as it happens
within a classical context, make subjective also the physical
reality itself. All these facts seem to find a confirmation in
some statements by Griffiths himself \cite{ref6}:
\begin{quotation}
{\small A classical analogy, that of ``coarse--graining'' in
classical statistical mechanics, is helpful in seeing why the
physicist`s freedom in choosing a quantum framework does not make
quantum theory subjective, or imply that this choice influences
physical reality. Coarse graining means dividing the classical
phase--space into a series of cells of finite volume. From the
point of view of classical mechanics, such a coarse--graining is,
of course, arbitrary; cells are chosen because they are convenient
for discussing certain problems, such as macroscopic
(thermodynamic) irreversibility. But this does not make classical
statistical mechanics a subjective theory. And, in addition, no
one would ever suppose that by choosing a particular
coarse-graining, the theoretical physicist is somehow influencing
the system.}
\end{quotation}

As we have claimed various times, within Classical Mechanics
there are two levels of reasoning: the first one makes reference to our
knowledge about a physical system, and the second one
to the properties objectively possessed
by the system and to their reality, quite independently from our
knowledge about them. With reference  to the previously analyzed example of
tossing a coin we can state:

\begin{enumerate}
    \item[a)] Our {\bf knowledge} about the outcome is that there is a
probability
  $1/2$ of getting {\it heads} and a probability  $1/2$ of getting {\it tails}:
this is all what we know.
    \item[b)] However, we can also argue in terms of  {\bf objectively
possessed
properties}, and we can legitimately claim that the outcome will surely be
either {\it heads} or {\it tails} and we can also claim that in the case in
which
the outcome will be {\it heads}, then certainly it will not be {\it tails}
 (this being simply a more lengthy way to state that the truth values
which are attached to all possible events concerning the tossing of a coin
form a homomorphism).  These statements might seem rather trivial and
uninteresting, but
the perspective changes radically, as implied by the  Kochen and  Specker
theorem,
when one takes into account that they refer
not to classical but to quantum properties.

\end{enumerate}
Accordingly, if one takes this position about the  ``single family'' rule,
the truth value of the decoherent histories is determined once for all and it
does not depend on the particular family we use to describe such
properties. It is only our (probabilistic) way of reasoning which must be
confined to a single family of decoherent histories. This means that it is
possible to choose the homomorphisms associated to the various families in
such a way that:
\begin{equation} \label{5.2}
    h_{\tiny\fam}(\hisa) = \makebox{const.},\qquad \forall\,\makebox{
    decoherent \fam},\,\,\hisa\in\fam,
\end{equation}
In our opinion, this is the only reasonable way to interpret the  ``single
family rule''.

\subsection{More about the ``single family rule''.}

    At this point, it could be useful to stress the basic difference between
the problem discussed in the example of section 3.3, which is
considered as irrelevant by the supporters of the DH, and the question raised
in the previous paragraph, i.e. whether or not the truth value of a given
history depends on the family to which it belongs. In the above mentioned
example, one tries to derive a contradiction by taking into account the
conjunction of  {\it different histories} belonging to different and
incompatible families; on the contrary, in the previous paragraph we have
considered {\it the same history} even though we view it as a member of
different decoherent families. Moreover,  the argument of the
example is entirely based on the use of the logical implication
``$\Longrightarrow$'' and, consequently, it deals with the {\it
probabilities} which characterize the various decoherent histories, with
the (probabilistic) {\it knowledge} we have about the system and its properties.
On the contrary, the arguments of the previous section call into play only the
{\it truth values} of the histories: in other words,  we have argued
only with reference to the  {\it objective properties}
possessed by physical systems, quite independently from our knowledge about
them.

    We can make more understandable the difference between the two
arguments  by resorting to the logical analysis of DH. Fundamentally,
following the previous example,  one would be tempted to claim that:
\begin{equation}
\label{5.3}
    \left\{\hisa\,\Longrightarrow\,\hisb,\,\,\,\hisa\,\Longrightarrow\,
    \hisg\right\} \supset \left\{\hisa\,\Longrightarrow \left[\hisb \wedge
    \hisg\right]\right\}.
\end{equation}
However, since $\hisb\wedge\hisg$ {\it does not belong to any decoherent
family}, it is not given any probability to it.
Correspondingly the logical implication:
\begin{equation}
    \hisa\,\Longrightarrow \left[\hisb
    \wedge\hisg\right].
\end{equation}
cannot even be defined. In our opinion this way of interpreting the ``single
family rule'' is the only sensible one which can be accepted.
    On the other hand, in the previous section we have asked whether, since the
{\it same}  history can belong to different decoherent families and if in all
such families it has a precise truth value:
\begin{eqnarray} \label{5.5}
    & &\left\{\hisa\in\makebox{Decoherent
    $\fam_{1}$}\right\}\supset\exists\,
    h_{1}\, :\, h_{1}[\hisa] \Longrightarrow \{0,1\},\nonumber \\
    & &\left\{\hisa\in\makebox{Decoherent
    $\fam_{2}$}\right\}\supset\exists\,
    h_{2}\, :\, h_{2}[\hisa] \Longrightarrow\{0,1\},
\end{eqnarray}
one should require that:
\begin{equation}
\label{5.6}
    h_{1}[\hisa]\quad =\quad h_{2}[\hisa].
\end{equation}
In our opinion, in no paper proposing the DH interpretation of Quantum
mechanics a clear cut answer to this fundamental question has been
given. In its formulation the ``single family rule'' is mute about this
problem. This is why we have considered various possible alternative readings
of the rule and we have called attention to their conceptual implications.

\section{Decoherent histories and the Kochen and Specker theorem.}

    In previous Sections we have analyzed some  points of the DH
approach which seem to us not to have been adequately clarified
and which make to some extent problematic the interpretation of
the theory. We have suggested what we consider the most obvious
and natural position about it if it has to be taken as a serious
alternative to the orthodox interpretation allowing to take a
realistic position about natural processes, the principal aim of
the proponents of this scheme.  In particular we stress the
following points:

\begin{enumerate}
\item The probabilities which the theory attaches to decoherent
histories should have the same ontological status as the classical
probabilities,
i.e. they should  refer to objective properties which may or may not be
actually possessed by the physical system we are dealing with.
Accordingly, for every decoherent family there should be an homomorphism from
the
Boolean set of its histories onto the Boolean set $\{0,1\}$:
    \[ \makebox{$\fam$ is Decoherent}
    \Rightarrow \exists \,\,h_{\tiny\fam}:\fam\rightarrow \{0,1
    \} \]

\item The homomorphism under 1) should depend on the history but not on the
different decoherent families to which it may belong:
    \[ h_{\tiny\fam}(\hisa) = \makebox{const.} =
    h(\hisa),\,\,\forall\,\makebox{Decoherent $\fam$ such that}
    \,\hisa\in\fam, \]
i.e., the truth value of a history and consequently also the physical
properties of which such history speaks,
should be independent from the particular
decoherent family one is dealing with. Here, then,
we interpret the ``single family''
rule as referring only to our knowledge of physical systems, not to their
properties, as we have discussed in section 6.
\end{enumerate}
In the following we shall prove that these two requirements cannot hold
simultaneously, unless one drastically restricts the set of allowed
Decoherent Families.

\subsection{The theorem.}

Consider the following four assumptions:
\begin{enumerate}
\item[a)] Every decoherent family has a natural Boolean structure which, for
one--time families, is defined as in section 3.1.

\item[b)] For every decoherent family there is an homomorphism from the
Boolean set of its histories onto the Boolean set $\{0,1\}$:
    \[ \makebox{$\fam$ is Decoherent}
    \Rightarrow \exists \,\,h_{\tiny\fam}:\fam\rightarrow \{0,1
    \}. \]

\item[c)] The homomorphism under b) depends on the history
but not on the
different decoherent families to which it may belong:
    \[ h_{\tiny\fam}(\hisa) = \makebox{const.} =
    h(\hisa),\,\,\forall\,\makebox{Decoherent $\fam$ such that}
    \,\hisa\in\fam. \]

\item[d)] All decoherent families have the same
status and must be taken into account;
\end{enumerate}
We now prove that they lead to a contradiction. Before proceeding
we would like to stress that what we are going to derive is a
mathematical theorem which is logically implied by the just
mentioned assumptions\footnote{Even though the above assumptions
have implications concerning the meaning of the theory, they are
formulated in a precise way. Therefore, if somebody, like
Griffiths and the referee, believe that we are violating the
``single family rule'', he has to tell clearly which one (or ones)
of the assumptions violates such a rule. After having shown that
such a rule is violated and having identified the precise
assumption which he thinks clashes with the rule, he should
plainly declare that he accepts the consequences of his choice,
consequences which we have analyzed in details in the previous
Sections.}.

    Consider a set $\dhis$ of histories (which belong to
decoherent families) involving only one time instant\footnote{We recall
that a one--time family contains histories of the type $(P_{k},t)$,
$t$ being an arbitrary time instant and the operators $\{P_{k}\}$ being an
exhaustive and exclusive set summing up to the identity operator. Note
that every one--time family is automatically decoherent.} $t$:
\begin{equation} \label{7.3}
    \dhis = \left\{\His = (P,t),\,\,\, \makebox{$P$ is a projector
    operator, $t$ is a fixed time}\right\}.
\end{equation}
Recall that $\dhis$  contains the null history  $(0,t)$
and the identity history  $(1,t)$.

    We now remark that, taking advantage of the Boolean character
of every family, the set $\dhis$ can be naturally equipped with the
structure of a partial Boolean algebra (PBA). A PBA is a set $B$  with
two distinguished elements $0$ and $1$, the meet $(\vee)$,
join $(\wedge)$ and complement $({}^{\perp})$ relations and
a commensurability relation $R$  satisfying the following axioms:
\begin{eqnarray} \label{7.4}
    & 1. & \forall\, x \in B:\,\,\, x R x. \nonumber \\
    & 2. & \forall\, x,y \in B:\,\,\, x R y \Rightarrow y R x. \nonumber \\
    & 3. & \forall\, x \in B:\,\,\, 0 R x,\,\, 1 R x. \nonumber \\
    & 4. & \forall\, x,y \in B:\,\,\, x R y \Rightarrow x R y^{\perp}.
        \nonumber \\
    & 5. & \forall\, x,y,z \in B: (x R y,\,\, x R z,\,\, y R z) \Rightarrow
        x R (y \vee z),\,\, x R (y \wedge z). \\
    & 6. & \forall\, x \in B:\,\,\, x \vee x = x. \nonumber \\
    & 7. & \forall\, x \in B:\,\,\, 0 \vee x = x \vee 0 = x,\,\, 1 \wedge x
        = x \wedge 1 = x. \nonumber \\
    & 8. & \forall\, x \in B:\,\,\, x \wedge x^{\perp} = 0,\,\, x \vee
            x^{\perp} = 1. \nonumber \\
    & 9. & \forall\, x,y,x \in B:\,\,\, (x R y,\,\, x R z,\,\, y R z)
        \Rightarrow \nonumber \\
         & & \Rightarrow x \wedge (y \vee z) = (x \wedge y) \vee (x
         \wedge z),\,\, x \vee (y \wedge z) = (x \vee y) \wedge (x \vee
         z). \nonumber
\end{eqnarray}

    The commensurability relation can be easily defined: we say
that two histories $\His^{(1)}$  and $\His^{(2)}$ are commensurable {\bf iff}
they are compatible, i.e. they belong together to at least one decoherent
family. Note that since in our  proof we are considering only
one--event histories, compatibility reduces to commutativity of the
projection operators appearing in the two histories:
\begin{equation} \label{7.5}
    \His^{(1)} = (P_{1},t);\,\,\, \His^{(2)} =
    (P_{2},t);\quad \His^{(1)}\, R\,
    \His^{(2)}\,\Leftrightarrow [P_{1}, P_{2}] = 0.
\end{equation}
Moreover, the meet, join and complement operations are defined like
in Section 3. It is now immediate to check that if  $[P_{1}, P_{2}] = 0$
then the first 5 axioms (\ref{7.4}) are satisfied. Also the
remaining axioms are easily checked to hold. Just to give an example we
remark that the last axiom requires the distributivity property to be
satisfied only for triplets of elements which are commensurable with each
other, i.e. for triplets of histories belonging to the same decoherent
family. On the other hand we know that in a decoherent family (due to its
Boolean structure) the distributive property holds.

    The second step is to remark that due to the fact that for any
decoherent family there exists a two--valued homomorphism
(assumption b) such that the image does not depend on the family
(assumption c), we can also define in $\dhis$  a two--valued
homomorphism  $H$. We recall that a homomorphism $h$ between two
PBAs, $B$  and  $C$ is a relation satisfying the following properties:
\begin{eqnarray}
    & 1. & \forall\, x,y \in B:\,\,\, x R y\,\Rightarrow\, h(x) R h(y).
        \label{7.6a} \\
    & 2. & \forall\, x \in B:\,\,\, h(x^{\perp}) = [h(x)]^{\perp}.
    \label{7.6b} \\
    &  3. & \forall\, x,y \in B: x R y\,\Rightarrow\, h(x \wedge y) = h(x)
        \wedge h(y),\,\, h(x \vee y) = h(x) \vee h(y). \label{7.6c}
\end{eqnarray}

    Now we can go on by defining the homomorphism $H$ according to:
\begin{equation} \label{7.7}
    H:\,\dhis\,\rightarrow\,\{0,1\}\,\,\,\makebox{by putting}\,\,\,
    H(\His) = h(\His),
\end{equation}
where $h(\His)$ is the homomorphism discussed in Section 5 and 6. It is
immediate to check that $H$  has all the required properties. It is
important to stress that $H$  is unambiguously and correctly defined just
because the homomorphism $h$ depends only on the histories and not on
the families, i.e., any history has a precise truth--value, $0$ or $1$. If
$h(\His)$ were family--dependent it would not be guaranteed that one could
define $H$  for the one time histories.

Now we remark that $\dhis$  is isomorphic to $P(\mathcal{H})$ , the set of
all projection operators of the Hilbert space  $\mathcal{H}$ (which, in
turn, is a PBA, the commensurability relation being, once more,
commutativity between projection operators). The isomorphism is obviously
given by the correspondence:
\begin{eqnarray} \label{7.8}
    & & \Phi:\, \dhis\,\rightarrow\, P(\mathcal{H}) \nonumber \\
    & & \His \equiv (P,t)\,\Leftrightarrow\,\Phi(\His) = P.
\end{eqnarray}
Once we have shown that $\dhis$  and  $P(\mathcal{H})$  are isomorphic, we can
``carry'' the two--valued homomorphism $H$  from $\dhis$
to  $P(\mathcal{H})$ as follows\footnote{It is immediately checked that $K$
is a homomorphism.}:
\begin{equation} \label{7.9}
    K:\, P({\mathcal H})\,\rightarrow\,\left\{ 0, 1\right\},\quad
    P\,\rightarrow H[\Phi^{-1}(P)].
\end{equation}

    Let us summarize the situation. We have proved that if
conditions a)--c) hold, then one can define a homomorphism between the
PBA $P(\mathcal{H})$  of the projection operators on $\mathcal{H}$  which
are associated to histories of the set $\dhis$  and the Boolean set
$\{0,1\}$. Assumption d) of the theorem amounts to claim that the just
considered PBA includes all projections operators on  $\mathcal{H}$. But
Kochen and Specker, in their celebrated paper \cite{ref23}, have proved
precisely that such a homomorphism cannot exist. Thus we have
proved\footnote{We mention
that a generic --- taking few lines ---  statement about the possibility of
working out such a proof appeared in a paper by S. Goldstein and D.N. Page
\cite{ref17}.} that one cannot add to assumptions a) -- c) we have put forward
at the beginning of this Section, and which according to us are necessary
ingredients  to make the decoherent histories approach physically sensible,
the request that all decoherent families have to be taken into account without
meeting a contradiction.

\subsection{A clarifying example.}

    We try to
further clarify the situation by discussing the ``paradigmatic''
example of a spin 1 particle for which we limit our considerations
only to the spin degrees of freedom; another example has appeared
in ref. \cite{baghi}. Be $\sqx, \sqy, \sqz$ the squares (in units
of $\hbar^{2}$) of the spin components along three orthogonal
directions $x, y$ and $z$, respectively. The three considered
operators commute and have a complete set of common eigenvectors
which we will denote by $|\alpha\rangle$, $|\beta\rangle$ and
$|\gamma\rangle$, satisfying:
\begin{center}
\begin{tabular}{ccc}
    & & \\ \quad $\sqx |\alpha\rangle = |\alpha\rangle$ \quad & \quad
    $\sqx |\beta\rangle = |\beta\rangle$ \quad & \quad
    $\sqx |\gamma\rangle = 0$ \quad \\
    & & \\
    \quad $\sqy |\alpha\rangle = |\alpha\rangle$ \quad & \quad
    $\sqy |\beta\rangle = 0$ \quad & \quad
    $\sqy |\gamma\rangle = |\gamma\rangle$ \quad  \\
    & & \\
    \quad $\sqz |\alpha\rangle = 0$ \quad & \quad
    $\sqz |\beta\rangle = |\beta\rangle$ \quad & \quad
    $\sqy |\gamma\rangle = |\gamma\rangle$\quad \\
    & &  \\
\end{tabular}
\end{center}
If we consider the projection operators on the one--dimensional
manifolds spanned by the above states $(\pa = |\alpha\rangle\langle\alpha|,
\pb = |\beta\rangle\langle\beta|$  and $\pg = |\gamma\rangle\langle\gamma|)$
we have:
\begin{equation} \label{7.11}
    \sqx = \pa + \pb;\quad \sqy = \pa + \pg;\quad \sqz = \pb + \pg.
\end{equation}
Let us now consider the one--time family $\fam^{(x,y,z)}$
whose maximally fine-grained histories are characterized
by the projection operators $\{\pau\} \equiv \{\pa, \pb, \pg\}$, i.e. the
histories:
\begin{equation} \label{7.12}
    \hisa = \{\pa, \tu\};\quad \hisb = \{\pb, \tu\};\quad \hisg = \{\pg,
    \tu\},
\end{equation}
where $\tu$  is a time instant following the initial time $t_{0}$ in which
the spin state of the particle is, as usual, described by the statistical
operator $W$.

    We note that  $\fam^{(x,y,z)}$ is automatically a decoherent family and
that only one of the three considered histories $\hisa, \hisb$ and $\hisg$
is true. This is due to the fact that the considered histories are mutually
exclusive:
\begin{equation} \label{7.13}
    \hisa\wedge\hisb = \hisa\wedge\hisg = \hisb\wedge\hisg = 0,
\end{equation}
and exhaustive:
\begin{equation} \label{7.14}
    \hisa\vee\hisb\vee\hisg = 1.
\end{equation}

    Let us suppose now that $\hisa$  is true. Then:
\begin{equation} \label{7.15}
    h[\hisa] = 1;\quad h[\hisb] = 0;\quad h[\hisg] = 0.
\end{equation}
This means that:
\begin{eqnarray} \label{7.16}
    & & h[\sqx] = h[\hisa\vee\hisb] = h[\hisa]\vee h[\hisb] = 1 \nonumber \\
    & & h[\sqy] = h[\hisa\vee\hisg] = h[\hisa]\vee h[\hisg] = 1 \\
    & & h[\sqz] = h[\hisb\vee\hisg] = h[\hisb]\vee h[\hisg] = 0 \nonumber
\end{eqnarray}
We can now choose a new set of three orthogonal axes by going from $(x, y, z)$
to $(x', y', z)$ and we consider the new one--time decoherent
family  $\fam^{(x',y',z)}$ whose maximally fine--grained histories are
associated to the projection operators $\{P^{\alpha'_{1}}_{1}\} = \{\pap, \pbp,
\pgp\}$, which are such that:
\begin{equation} \label{7.18}
    \sqxp = \pap + \pbp;\quad \sqyp = \pap + \pgp;\quad \sqzp = \pbp + \pgp
    \equiv \sqz.
\end{equation}
In the above equation $\sqxp, \sqyp, \sqz$ are, as usual, the square of the spin
components along the indicated directions. Now, since $\pb + \pg = \sqz = \pbp +
\pgp$, the two coarse-grained histories $\His^{(\beta + \gamma)} \equiv \{\pb +
\pg, \tu\} \in \fam^{(x,y,z)}$ and $\His^{(\beta' + \gamma')} \equiv \{\pbp +
\pgp, \tu\} \in \fam^{(x',y',z)}$ are, de facto, the same history.
Due to the validity of assumption c), we can state that:
\begin{equation} \label{7.19}
    h[\His^{(\beta')}]\vee h[\His^{(\gamma')}] = h[\His^{(\beta' +
    \gamma')}] = h[\His^{(\beta + \gamma)}] = h[\His^{(\beta)}]\vee
    h[\His^{(\gamma)}] = 0,
\end{equation}
i.e., since one and only one of the three maximally fine-grained
histories must be true:
\begin{equation} \label{7.20}
    h[\His^{(\beta')}] = h[\His^{(\gamma')}] = 0;\quad
    h[\His^{(\alpha')}] = 1.
\end{equation}
We have thus shown that:
\begin{equation}
    h[\sqzp] = h[\sqyp] = 1;\quad h[\sqzp] = 0.
\end{equation}
The argument can obviously be repeated for any pair of orthogonal
triples having an axis in common. But Kochen and Specker have proved
that when the game is played for more than 117 appropriately chosen
triples one gets a contradiction.

    Taking the risk of being pedantic we want to stress once more
that our proof never requires to take into account different
histories belonging to incompatible decoherent families. The key
ingredient of the proof is the assumption that the same history
(in the above example $\His^{(\beta + \gamma)} \equiv
\His^{(\beta' + \gamma')}$), even when considered as belonging to
incompatible decoherent families ($\fam^{(x,y,z)}$ and
$\fam^{(x',y',z)}$), has the same truth value.

\section{What about IGUSes which can communicate?}

    Gell-Mann
and Hartle have also put forward \cite{ref25} the hypothesis that different
IGUSes belonging to different --- and in general incompatible ---
``quasi--classical domains'' may communicate with each other:
\begin{quotation}
{\small If different realms [i.e. quasi--classical domains]
exhibit IGUSes, we may investigate certain relations between them.
Probabilistic predictions concerning the relationships between
IGUSes in two different realms may be made by using a decohering
set of histories containing alternatives referring to IGUSes in
one realm and also alternatives referring to IGUSes in the other
realm, provided the decoherence of the hybrid set follows from the
initial conditions and Hamiltonian. The problem of drawing
inferences in one realm concerning IGUSes using a distinct realm
is then not so very different from that involved in ordinary
searching for extraterrestrial intelligence. The IGUSes making use
of one realm could conceivably draw inferences about IGUSes in
another by seeking or creating ``measurement situations'' in which
an alternative of one realm is correlated almost perfectly with an
alternative from the other.}
\end{quotation}

    Before proceeding a specification is necessary. In the above sentence
reference is made to ``different realms'' and to a ``hybrid set''
satisfying the decoherence  conditions containing alternatives
referring to two realms. In what follows we will interpret the
above statements as referring to the situation in which
consideration is given to incompatible families (different realms)
such that some (but not all)  histories belonging to different
families (the hybrid set) can be accommodated in a decoherent
family. The only other alternative, i.e., that the different
realms and the hybrid set are members of a unique decoherent
family would render the argument meaningless.

Having made clear this point  we remark that in the above sentence the
 authors tacitly assume that the {\it alternatives referring to IGUSes in
one realm and also alternatives referring to IGUSes in the other
realm} must possess the same truth value, both if they belong to
the first realm or to the second one, obviously given for granted
that in each realm and in the hybrid set the decoherence condition
be satisfied. Actually, only if this happens the two IGUSes can be
sure to be able to communicate correctly between them. More
generally we stress that the very way Gell--Mann and Hartle
contemplate as the mean for the two IGUSes to exchange information
coincides with the way used by us to relate the spin properties
referring to two different triplets of orthogonal directions with
one direction in common: the two triplets are associated to two
different --- and incompatible --- families of decoherent
histories having a common coarse--graining, the history referring
to the square of the spin along the axis they have in common. In
Gell--Mann and Hartle's example the same happens, and it is just
due to the common coarse--graining that the two IGUSes can
exchange information. Thus, in a certain sense, the argument of
these authors represents a further proof of the reasonableness and
naturalness of the assumptions we have made to prove the theorem
of Subsection 7.1.

    But if one takes into account our conclusion, one realizes that
the communication between IGUSes could turn out to be rather
problematic. In fact, suppose that two IGUSes, which we will call
IGUS1 and IGUS2, belong to realm 1 and realm 2, respectively, and
these realms have a common coarse graining which can be used for
exchanging information between them. Now one can consider IGUS3,
belonging to realm 3, having a coarse graining in common with
IGUS2, so that also these two IGUSes can communicate between them.
We can go on assuming the existence of further IGUSes, each in a
different realm, having a coarse graining in common with the one
preceding it. The realm of the last of these additional IGUSes,
IGUSN is assumed to have a coarse graining in common with the
realm of IGUS1. Now IGUS1 can exchange information with IGUS2
which in turn can transfer it to IGUS3 and so on up to the moment
in which the information reaches IGUSN which transmits this
information to IGUS1. The example with the spin components
analyzed in Subsection 7.2 shows that the information that is
transmitted to IGUS1 may contradict the properties  which are part
of the information already possessed by IGUS1.

    We are aware that this example looks more like a science--fiction
story than as a physical compelling argument, but we have considered
appropriate to discuss it for its implications concerning those attempts which
take the DH approach as the appropriate one for dealing with a framework in
which all statements about physical processes, properties possessed by physical
systems, and so on, are reduced to information exchange between IGUSes.

\section{Summary, conclusions and perspectives.}

    We close this long paper by summarizing the fundamental questions
regarding the interpretation of the  DH approach which have not yet received
(in our opinion) a clear cut answer, by stating once more which assumption we
consider necessary and by suggesting the line to follow to work out a
satisfactory theory.
\begin{itemize}
    \item  The probabilities assigned to decoherent histories are
    {\it probabilities of what, exactly}? The two possible alternatives are:

\begin{enumerate}
    \item Just as in the case of Classical Mechanics {\it the probabilities
associated to decoherent histories refer to objective properties
of individual physical systems}. This implies that they must have a truth
value and, as discussed in great detail, this in turn can be formalized by
considering an appropriate homomorphism between the histories of a decoherent
family and the set $\{0,1\}$. This seems to us the only reasonable solution if
one pretends that the DH approach represents an actual improvement of the
standard interpretation allowing to take a realistic attitude with respect to
physical processes.
    \item {\it The DH probabilities (or at least some of them) do not refer to
objective properties of physical systems}. If this is the case the
proponents of the DH approach should make clear `what they are
probabilities of' and should put forward precise criteria to
identify those histories which have a truth value. They should
also make clear why these histories have a privileged status.
Griffiths' proposal to attach truth values {\it only} to histories
having probability equal to one or zero (if this is what he means)
is useless. In our opinion no reasonable solution to the
fundamental problems of quantum mechanics can be reached following
such a line.
\end{enumerate}

    \item The ``single family rule'' applies only to our knowledge about the
properties of physical systems and to the way we can use such information to
make (probabilistic) logical inferences or does it hold also for the
properties themselves?
\begin{enumerate}

    \item If {\it the ``single family'' rule concerns only our knowledge} and
not the actual properties of physical systems which have a truth value
completely independent from our knowledge, then such truth values
cannot change by changing the decoherent family. In our language, the
homomorphism is family independent. In our opinion this is the only
reasonable attitude about this point.
\item If {\it the rule applies also to the properties of physical systems},
and, consequently the truth value of a history depends on the decoherent
family to which it is considered to belong (in the precise sense that the
truth value can change with the family) then there follows that the properties
of which the theory speaks are not objective but they depend in a crucial way
from the framework one chooses to describe the system. Such a situation is, in
our opinion, unacceptable.
\end{enumerate}
\end{itemize}

As we have openly stated, we believe that the choices labeled by 1
must be made for both questions we have raised. The reasons for
this choices have been analyzed in detail and we urge the readers
to take into account the precise ontological implications of
dropping any one of them. As repeatedly remarked, we believe that
any such choice  renders the DH view at least as problematic as
the standard interpretation of Quantum Mechanics.

Now we can analyze the final and crucial question we have repeatedly mentioned:
\begin{itemize}
    \item Do all decoherent histories have the same ontological status, and, in
particular, do they speak of properties objectively possessed by individual
physical systems? Two answers are possible:

\begin{enumerate}
    \item Yes, {\it all decoherent histories make reference to objective
properties and have definite truth values}. If this is the case it
is totally obscure how decoherent histories referring to
physically unacceptable properties for macroscopic systems (such
as  their being in a superposition of macroscopically different
states) can have any sensible physical meaning. Moreover we stress
that the theorem of Section 7 shows that this choice is
incompatible with  the choices we have made concerning  both
previous questions. Since  we consider such choices as necessary
to have a meaningful theory which represents a real improvement on
the standard interpretation of the theory, we are led to consider
the alternative answer to the above question.

    \item {\it Not all decoherent histories have a physical meaning}, only those
belonging to an appropriate subset are associated to objective
properties of physical systems, and, consequently, must have a
truth value. In such a case  non--ambiguous criteria to identify
the histories which have truth values must be put forward. With
reference to this point we remark that:

\begin{enumerate}
    \item {\it Omn\`es'} criterion to distinguish  sensible
families
    from senseless ones (which looks quite promising) does not work.
    \item  {\it Griffiths'} criterion (when correctly interpreted) to attach
truth values only to histories having probability one or zero is senseless
since no decoherent realistic history can have such a probability of
occurrence.
    \item The  {\it Gell--Mann and Hartle}'s suggestion to introduce a measure 
of
classicity to characterize the physically significative histories is
extremely interesting but, up to now, it has not found a satisfactory
formulation.
\end{enumerate}
\end{enumerate}
\end{itemize}

Concluding, the just indicated line seems to us the only
interesting one: one must drastically reduce, by precise criteria,
the set of decoherent families which can be considered and which
are physically significative. Obviously, if one takes such  an
attitude the real problem is to work out a consistent criterion to
identify the acceptable families.

\section*{Acknowledgments}
We acknowledge useful remarks and suggestions by the referee.

\end{document}